\documentclass[12pt]{article}
\begin{document}

\arraycolsep 0pt
\newcommand{\be}{\begin{equation}}
\newcommand{\ee}{\end{equation}}
\newcommand{\ba}{\begin{eqnarray}}
\newcommand{\ea}{\end{eqnarray}}
\newcommand{\dalam}{\raisebox{1mm}{\fbox{}{}}\;}
\newcommand{\pa}{\partial}
\newcommand{\f}{\frac}
\newcommand{\st}{\stackrel}
\newcommand{\tk}{\tilde\kappa}
\newcommand{\ep}{\epsilon}

\author{A.A.Logunov, M.A.Mestvirishvili\\{\it Institute for High
Energy Physics,}\\
{\it 142284, Protvino, Moscow region, Russia}}
\title{\Large  \bf WHAT HAPPENS IN THE VICINITY OF THE SCHWARZSCHILD
SPHERE WHEN
NONZERO
GRAVITON REST MASS IS PRESENT}
\date{}
\maketitle

\begin{abstract}
In this paper a solution for a static spherically symmetric body is
thoroughly considered
in the framework of the Relativistic Theory of Gravitation. By the
comparison
of this solution with the Schwarzschild solution in General
Relativity 
their substantial difference is established 
in the region close to the
Schwarzschild sphere. Just this difference excludes the possibility
of collapse
to form ``black holes''.
\end{abstract}

The given problem was considered for the first time  in the
Relativistic Theory of
Gravitation (RTG) in paper~[1], where it was established that in
vacuum the metric coefficient $g_{00}$ of the effective Riemannian
space
was not equal to zero on
the Schwarzschild sphere, whereas $g_{11}$ had a pole. These
changes which have arisen in the theory because of the graviton mass
result
in   a ``bounce'' effect of the falling particles and light from a
singularity
on the Schwarzschild spere, and consequently, in the absence of
``black
holes''.

Later in paper~[2] an in-depth study of this problem in
the RTG was conducted  which  updated a number of points, but at the
same time
showed, that  the ``bounce'' took place close to the
Schwarzschild sphere. In
view of importance of this  problem we again come back to its
analysis
with the purpose of showing in a simpler and clearer way that in that
point in vacuum
where the metric coefficient of effective Riemannian space $g_{11}$
has a pole,
another metric coefficient $g_{00}$ will not vanish.

In RTG [3] the gravitational field is considered as a physical field
in the Minkowski space. The source of this field is the universal
conserved density of the
energy-momentum tensor
of the entire matter including the gravitational
field. This
circumstance results in the emerging of the effective Riemannian
space
because of
the presence of the gravitational field. The motion of matter in the
Minkowski space
under the influence of the gravitational field proceeds in the same
way as if it moved in the
effective Riemannian space. The field approach to gravitation with
necessity
requires the introduction of the graviton rest mass.

In RTG, as opposed to the
General Relativity Theory (GRT), the inertial reference frames are
present and consequently
the acceleration has an absolute meaning. The forces of inertia and
gravity are
separated, as they are of completely different nature. The Special
Relativity Principle
holds  for all  the physical fields, including the gravitational one.
It follows from this theory that gravitational forces in the
Newtonian approximation are the forces of attraction. Since a
physical field
can be described
in one coordinate system, it means, that the effective Riemannian
space has a 
simple topology and is set in one chart. In RTG the Mach Principle
will be realised  --- an
inertial reference frame is determined by the
distribution of matter. In this theory the Correspondence Principle
takes
place: after switching off the gravitational field the curvature of
space disappears, and we
find ourselves in the Minkowski space in the coordinate system
prescribed earlier.

The RTG equations look like
\be
R^\mu_\nu -\f{1}{2}\delta^\mu_\nu R+\f{1}{2} \left (
\f{mc}{\hbar}
\right )^2 \left ( \delta^\mu_\nu +g^{\mu\alpha}
\gamma_{\alpha\nu}-\f{1}{2}\delta^\mu_\nu
 g^{\alpha\beta}\gamma_{\alpha\beta}
\right )
=\kappa T^\mu_\nu,
\ee
\be
D_\mu\tilde g^{\mu\nu}=0\;.
\ee
Here $ \tilde g^{\mu\nu} = \sqrt{-g} g^{\mu\nu}, \; g = \mbox{det} \;
g_{\mu\nu}, \; R^\mu_\nu $ is the Ricci tensor, $ \kappa = \f{8\pi
G}{c^2}, \; \; G $ is the gravitational constant, $D_\mu $ is the
covariant derivative in the Minkowski space, $ \gamma_{\mu\nu}(x) $
is the metric tensor of the Minkowski space in arbitrary curvilinear
coordinates. Equations (1) and (2) are covariant under arbitrary
coordinate transformations with a nonzero Jacobian. They are also
Lorentz invariant
under transformations from one inertial
system in Galilean coordinates to another. Equations (2)
eliminate   representations corresponding to
spins $1$ and $0'$ for a tensor field,  
leaving only the representations with spins 2 and
0. The equations of motion of matter are the consequents of equations
(1)
and (2).

Let us determine now the gravitational field created by a
spherically-symmetric static source. The general form of the interval
of the
effective Riemannian space for such source looks like
\be
ds^2=g_{00} dt^2+2g_{01}dt dr+
g_{11}dr^2+g_{22}d\Theta^2+g_{33}d\Phi^2.
\ee
Let us introduce the notations
\ba
g_{00}(r)=U(r),\; g_{01}(r)=B(r),\;
g_{11}(r)&=&
-\left [
V(r)-\f{B^2(r)}{U(r)}
\right ]\;,\nonumber \\[-0.2cm]
\\[-0.2cm]
g_{22}(r)=-W^2(r),\;
g_{33}(r,\Theta)&=&-W^2(r)\sin^2\Theta.\nonumber
 \ea
The components of the contravariant metric tensor are as follows:
\ba
g^{00}(r)=\frac{1}{U}
\left (
1-\f{B^2}{UV}
\right ),\;
g^{01}(r)&=&-\frac{B}{UV},\;
g^{11}(r)=-\frac{1}{V},\;\nonumber \\
[-0.2cm]
\\[-0.2cm]
g^{22}(r)=-\f{1}{W^2}\;,\;
g^{33}(r,\Theta)&=&-\f{1}{W^2\sin^2\Theta}.\nonumber
\ea
The determinant of the metric tensor $g_{\mu\nu} $ is equal to
\be
g=\mbox{det} g_{\mu\nu}=-UVW^4\sin^2\Theta\;.
\ee
For the solution having a physical sense, the following condition
should be satisfied:
\be
g<0\;.
\ee
For spherical coordinates  $g$ can be equal to zero only at a point
$r=0$. On the base of (5) and (6) we obtain the components of the
metric
tensor density
\be
\tilde g^{\mu\nu}=\sqrt{-g}g^{\mu\nu}.
\ee
They have the form 
\be
\tilde g^{00}=\f{W^2}{\sqrt{UV}} \left (
V-\f{B^2}{U}
\right ) \sin\Theta,\;
\tilde g^{01}=-\f{BW^2}{\sqrt{UV}} \sin\Theta,\;
\tilde g^{11}=-\sqrt{\f{U}{V}}W^2 \sin\Theta,\;
\ee

$$
\tilde g^{22}=-\sqrt{UV}\sin\Theta,\; \tilde
g^{33}=-\f{\sqrt{UV}}{\sin\Theta}\;.\eqno{(9')}
$$

All the consideration will be provided for an inertial system in
spherical
coordinates. The interval of the Minkowski space looks like
\be
d\sigma^2=dt^2-dr^2-r^2
(d\Theta^2+\sin^2\Theta d\Phi^2)\;.
\ee
Nonzero  Christoffel symbols of the  Minkowski space defined by the
following formula
\be
\gamma^\lambda_{\mu\nu}=\f{1}{2}
\gamma^{\lambda\sigma}
(\pa_\mu\gamma_{\sigma\nu}
+\pa_\nu\gamma_{\sigma\mu}-\pa_\sigma\gamma_{\mu\nu})\;
\ee
are equal to
\be
\gamma^1_{22}=-r,\;
\gamma^1_{33}=-r\sin^2\Theta,\;
\gamma^2_{12}=\gamma^3_{13}=\f{1}{r},\;
\gamma^2_{33}=-\sin\Theta\cos\Theta,\;
\gamma^3_{23}=\cot \Theta\;.
\ee

Let us write equations (2) in the extended form
\be
D_\mu\tilde g^{\mu\nu}=\pa_\mu \tilde g^{\mu\nu}
+\gamma^\nu_{\lambda\sigma}\tilde g^{\lambda\sigma}=0\;.
\ee
In Galilean coordinates of the Minkowski space they look like
\be
\pa_\mu\tilde g^{\mu\nu}=0\;.
\ee
In the case of a static gravitational field we have from (14)
\be
\pa_i\tilde g^{i\nu}=0,\;
i=1,2,3\;.
\ee

By using the tensor transformation law it is possible to express
components $\tilde g^{i0}$ in Cartesian coordinates through
components in spherical coordinates
\be
\tilde g^{i0}=-\f{BW^2}{\sqrt{UV}}\cdot
\f{x^i}{r^3}\;,
\sqrt{-g}=\f{1}{r^2}\sqrt{UV}W^2.
\ee
Here $x^i$ are spatial Cartesian coordinates. Supposing in (15)
$\nu=0$ and integrating over a spherical volume after applying
the Gauss-Ostrogradskii theorem, we get the integral over a spherical
surface
\be
\oint \tilde g^{i0}ds_i=
-\f{BW^2}{r^3\sqrt{UV}}\oint (\vec x  d \vec s)=0\;.
\ee
Taking into consideration the  equality
\be
\oint(\vec x d\vec s)=4\pi r^3,
\ee
we get
\be
\f{BW^2}{\sqrt{UV}}=0\;.
\ee
As  equation (14) is fair both inside matter, and outside of it
equation (19)
should be true for any value of $r$. But as due to eqution (7) $U,V$
and
$W$ cannot be equal to zero everywhere, it follows from (19) that
\be
B=0\;.
\ee
Interval (3) of the effective Riemannian spaces becomes
\be
ds^2=Udt^2-Vdr^2-W^2(d\Theta^2+\sin^2\Theta d\Phi^2)\;.
 \ee
From  equation (20) it follows, that there is no static
solution for the Hilbert-Einstein equations in harmonic coordinates
which would have in the
interval expression the term like
\be
B(r)dtdr\;.
\ee

The energy-momentum tensor of matter looks like
\be
T^\mu_\nu=\left (
\rho+\f{p}{c^2}\right ) v^\mu v_\nu-\delta^\mu_\nu\cdot \f{p}{c^2}\;.
\ee
In expression (23) $\rho$ is the mass density of matter, $p$ is the
isotropic pressure, and
\be
v^\mu=\f{dx^\mu}{ds}
\ee
is 4-velocity that  meets the condition
\be
g_{\mu\nu}v^\mu v^\nu=1\;.
\ee
From equations (1) and (2) it follows \be
\nabla_\mu T^\mu_\nu=0\;,
\ee
where $ \nabla_\mu $ is the covariant derivative in the effective
Riemannian
space with a metric tensor $g_{\mu\nu} $. In case of a static body
\be
v^i=0,\;\; i=1,2,3;\;\; v^0=\f{1}{\sqrt{U}}\;, \ee
and consequently
\be
T^0_0=\rho (r),\;\; T^1_1=T^2_2=T^3_3=
-\f{p(r)}{c^2}\;,\;\;
T^\mu_\nu=0,\;\; \mu\not =\nu\;.
\ee

For  interval (21)  the nonzero Christoffel symbols are
\ba
\Gamma^0_{01}=\f{1}{2U}\f{dU}{dr}\;,\;
\Gamma^1_{00}=\f{1}{2V}\f{dU}{dr}\;,\;
\Gamma^1_{11}&=&\f{1}{2V}\f{dV}{dr}\;,\;
\Gamma^1_{22}=-\f{W}{V}\f{dW}{dr},\nonumber \\ \\
[-0.2cm]
\Gamma^1_{33}=\sin^2\Theta\cdot \Gamma^1_{22}\;,\;
\Gamma^2_{12}=\Gamma^3_{13}=\f{1}{W}\f{dW}{dr}\;,\;\;
\Gamma^2_{33}&=&-\sin\Theta\cos\Theta\;,\;\;
\Gamma^3_{23}=\cot\Theta.\nonumber
 \ea
By using the following expression for the  Ricci  tensor
\be
R_{\mu\nu}=\pa_\sigma\Gamma^\sigma_{\mu\nu}
-\pa_\nu \Gamma^\sigma_{\mu\sigma}
+ \Gamma^\sigma_{\mu\nu}\Gamma^\lambda_{\sigma\lambda}
- \Gamma^\sigma_{\mu\lambda}\Gamma^\lambda_{\sigma\nu}\;,\;
R^\mu_\nu=g^{\mu\lambda}R_{\lambda\nu}
\ee
and substituting into it the expressions for
 the Christoffel symbols from (29), it is possible to reduce
 equations (1) for functions $U,V$ and $W$ to the following form: 
 \ba
\f{1}{W^2}&-&\f{1}{VW^2}
\left(
\f{dW}{dr}\right )^2
-\f{2}{VW}\f{d^2W}{dr^2}
-\f{1}{W}\f{dW}{dr}\f{d}{dr}
\left(
\f{1}{V}\right )+\nonumber \\[-0.2cm] \\ [-0.2cm]
&+&\f{1}{2}
\left(
\f{mc}{\hbar}\right )^2 \left[
1+\f{1}{2}
\left (
\f{1}{U}-\f{1}{V}\right )
-\f{r^2}{W^2}\right ]
=\kappa \rho\;,\nonumber
\ea
\ba
\f{1}{W^2}-\f{1}{VW^2}
\left (
\f{dW}{dr}\right )^2
-\f{1}{UVW}\f{dW}{dr}&\cdot& \f{dU}{dr}+\nonumber \\[-0.2cm] \\
[-0.2cm]
+\f{1}{2}
\left(
\f{mc}{\hbar}\right )^2 \left [
1-\f{1}{2}
\left (
\f{1}{U}-\f{1}{V}\right )
-\f{r^2}{W^2}\right ]
&=&-\kappa\f{p}{c^2}\;,\nonumber
\ea
\ba
-\f{1}{VW} W''-\f{1}{2UV}U''
&+&\f{1}{2WV^2}W'V'+\f{1}{4VU^2}
(U')^2+\nonumber \\[-0.2cm] \\ [-0.2cm]
+\f{1}{4UV^2}U'V'
-\f{1}{2UVW}W'U'
&+&\f{1}{2}
\left (
\f{mc}{\hbar}\right )^2 \left [
1-\f{1}{2}
\left (
\f{1}{U}+\f{1}{V}\right )\right ]=-\kappa\f{p}{c^2}\;.\nonumber \ea

Equation (13) after taking into account (12), (9) and (20) is
as follows: 
\be
\f{d}{dr}\left(
\sqrt{\f{U}{V}}W^2\right )=2r\sqrt{UV}\;.
\ee

Let us remark that by virtue of the  Bianchi identity and equation
(2) one
of
equations (31-33) is a consequent of the others. Further we shall
take equations (31), (32) and (34) as independent.

We shall write equation (26) in the extended form as
\be
\nabla_\mu T^\mu_\nu\equiv \pa_\mu T^\mu_\nu
+\Gamma^\mu_{\alpha\mu} T^\alpha_\nu
-\Gamma^\alpha_{\mu\nu}T^\mu_\alpha =0\;.
\ee
By using expressions (28) and (29) we  obtain
\be
\f{1}{c^2}\cdot\f{dp}{dr}=-\f{\rho+\f{p}{c^2}}{2U}
\cdot\f{dU}{dr}\;.
\ee
Taking into consideration  identity \be
\f{1}{W^2\left (\f{dW}{dr}\right )}\cdot \f{d}{dr} \cdot\left [
\f{W}{V} \left (
\f{dW}{dr}\right )^2\right]
=\f{1}{VW^2}
\left (
\f{dW}{dr}\right )^2
+\f{2}{VW}\cdot
\f{d^2W}{dr^2}
+\f{1}{W}\f{dW}{dr}\f{d}{dr}
\left (\f{1}{V}\right )\;, \ee
equation (31) can be written in the following form
\be
1-\f{d}{dW}
\left [
\f{W}{V\left (\f{dr}{dW}\right )^2}\right ]
+\f{1}{2}
\left (
\f{mc}{\hbar}\right )^2 \left [
W^2-r^2+\f{W^2}{2}
\left (
\f{1}{U}-\f{1}{V}\right )\right ]
=\kappa W^2\rho\;.
\ee
Similarly we transform equation (32): \be
1-\f{W}{V\left (\f{dr}{dW}\right )^2} \cdot \f{d}{dW}\ln
(UW)+\f{1}{2}\left (
\f{mc}{\hbar}\right )^2 \left [
W^2-r^2-\f{1}{2}
\left (\f{1}{U}-\f{1}{V}\right )\right ]
=-\kappa\f{W^2p}{c^2}\;.
\ee
We shall write Eqs. (34) and (36) as follows:
\be
\f{d}{dW}\left (W^2\sqrt{\f{U}{V}}\right )
=2r\sqrt{UV}\f{dr}{dW}\;.
\ee
\be
\f{1}{c^2}\cdot \f{dp}{dW}= -\left (\rho +\f{p}{c^2}\right )\f{1}{2U}
\cdot
\f{dU}{dW}\;.
\ee
Let us proceed to dimensionless variables in equations (38) -- (41).
Let $l$ be the Schwarzshild radius of the source which has  mass $M$
\be
l=\f{2GM}{c^2}\;.
\ee
Let us introduce new variables $x$ and $z$ which are equal to
\be
W=lx,\;\; r=lz.
\ee
Equations (38-41) become
$$
1-\f{d}{dx}
\left (
\f{x}{V\left (\f{dz}{dx}\right )^2} \right )
+\epsilon \left [
x^2-z^2+\f{1}{2}x^2
\left (
\f{1}{U}-\f{1}{V} \right )\right ]
=\tilde \kappa x^2\rho (x), \eqno{(38')}
$$
$$
1-\f{x}{V\left (\f{dz}{dx}\right )^2} \f{d}{dx}\ln (xU)+\epsilon
\left [
x^2-z^2-\f{x^2}{2}\left (
\f{1}{U}-\f{1}{V}\right )\right ]
=-\tilde \kappa \f{x^2p(x)}{c^2}\;, \eqno{(39')}
$$
$$
\f{d}{dx}
\left (
x^2\sqrt{\f{U}{V}}\right )
=2z\f{dz}{dx} \sqrt{UV}\;,\eqno{(40')}
$$
$$
\f{1}{c^2}\f{dp}{dx}=
\left (
\rho+\f{p}{c^2}\right )\f{1}{2U}\f{dU}{dx}\;.\eqno{(41')}
$$
Here $ \epsilon $ is a numerical constant which is equal to
\be
\epsilon =\f{1}{2} \left (
\f{2GMm}{\hbar c}\right )^2,\;\; \tilde \kappa =\kappa l^2.
\ee
The sum and difference of equations $(38')$ and $(39')$ are
\be
2-\f{d}{dx}
\left [
\f{x}{V\left (\f{dz}{dx}\right )^2} \right ]
- \f{x}{V\left (\f{dz}{dx}\right )^2}
\f{d}{dx}\ln (xU)+2\epsilon (x^2-z^2)
=\tilde \kappa x^2 \left (\rho-\f{p}{c^2}\right )\;,
\ee
\be
\f{d}{dx}
\left [
\f{x}{V\left (\f{dz}{dx}\right )^2} \right ]
-\f{x}{V\left (\f{dz}{dx}\right )^2}
\f{d}{dx}\ln (xU) -\epsilon x^2 \left (\f{1}{U}-\f{1}{V} \right )
=-\tilde \kappa x^2 \left (\rho +\f{p}{c^2}\right )\;.
\ee

Let us introduce new functions $A$ and $ \eta $: \be
U=\f{1}{x\eta A},\;\; V=\f{x}{A\left (\f{dz}{dx}\right )^2}\;.
\ee
In these new variables  equation (45) becomes \be
A\f{d\ln \eta}{dx}+2+2\epsilon (x^2-z^2)
=\tilde\kappa x^2 \left (\rho-\f{p}{c^2}\right )\;.
\ee
Equation $(38')$ can be written in the following form: \be
\f{dA}{dx}=1+\epsilon (x^2-z^2)
+\epsilon \f{x^2}{2} \left (
\f{1}{U} -\f{1}{V}\right ) -\tilde\kappa\cdot x^2\rho (x)\;.
\ee
According to the causality condition  (see ~ Appendix~L)
\be
\gamma_{\mu\nu}U^\mu U^\nu =0\;,
\ee
$$
g_{\mu\nu}U^\mu U^\nu \leq 0\;,\eqno{(50')}
$$
it is easy to establish the following inequality:
\be
U\leq V\;.
\ee
In our problem it is possible to limit ourselves by values $x $ and
$z $ from the following interval only: 
\be
0\leq x \ll \f{1}{\sqrt{2\epsilon}},\;\;
0\leq z \ll \f{1}{\sqrt{2\epsilon}}.\;\; \ee
These inequalities limit $r, W $ from above  by the value 
\be
r,W \ll \f{\hbar}{mc}\;.
\ee
Under such a limitation equation (49) becomes \be
\f{dA}{dx}=1+\epsilon \f{x^2}{2} \left (
\f{1}{U}-\f{1}{V}\right )-\tk x^2 \rho (x)\;.
\ee
Outside of the matter we have 
\be
\f{dA}{dx}=1+\epsilon\f{x^2}{2}
\left (
\f{1}{U}-\f{1}{V}\right )\;.
\ee
By virtue of the causality condition  (51) the following inequality
takes place outside of matter
\be
\f{dA}{dx} \geq 1.
\ee
Integrating (54) in an interval $ (0, x) $ we get
\be
A(x)=x+\f{\epsilon}{2}
\int\limits^{x}_{0}x^{'2}
\left (
\f{1}{U}-\f{1}{V}\right ) dx'-\tk
\int\limits^{x}_{0}x^{'2}\rho (x')dx'.
\ee
In (57) $A (0) $ is trusted to be equal to zero, as if it was
distinct
from zero, the function $V (x) $ would become zero when $x $ tends
to zero, that is physically forbidden. On the base of relation (56)
function $A
(x) $ monotonically increases with $x $ outside of matter, and
therefore it
can have only one root
\be
A(x_1)=0,\;\; x_1>x_0.
\ee
On the base of relation (57) we have \be
x_1=1-\f{\epsilon}{2}
\int\limits^{x_1}_{0}x^{'2}
\left (
\f{1}{U}-\f{1}{V}\right ) dx'.
\ee
Here we take into account that by selecting $l$ equal to (42)
$$
\tk
\int\limits^{x_0}_{0}x^{'2}
\rho (x')dx'=1.
$$
The matter is concentrated in the sphere  $0\leq x\leq x_0 $.

Because of a graviton mass, zero point of  function $A $ is shifted
inside the Schwarzshild sphere. As at $x $ tending to $x_1, \; \; V
(x) $ is tending to
infinity, since $A (x) $ is going to zero, there will be such
neighborhood of
a point $x_1 $
\be
x_1(1-\lambda_1)\leq x\leq x_1
(1+\lambda_2),\;\; \lambda_1>0, \lambda_2>0\;, \ee
( $ \lambda_1 $ and $ \lambda_2 $ receive small fixed values), in
which  the following
inequality  will take place
\be
\f{1}{U}\gg \f{1}{V}\;.
\ee
In this approximation we obtain
\be
A(x)=x-x_1+\f{\epsilon}{2}
\int\limits^{x}_{x_1}dx'x^{'2}\f{1}{U}\;.
\ee
Substituting  $U $ in the form given by relation
(47) into this expression, we shall discover
\be
A(x)=x-x_1+\f{\epsilon}{2}
\int\limits^{x}_{x_1}dx'x^{'3}\eta (x')A(x').
\ee
If a range of $x $ is in interval (60), then in the integrand it is
possible to change
$x^3 $ for $x^3_1 $: \be
A(x)=x-x_1+\f{\epsilon}{2}x^3_1
\int\limits^{x}_{x_1}\eta (x')A(x')dx'.
\ee
From here we get \be
\f{dA}{dx}=1+\f{\epsilon}{2}x^3_1\eta (x)A(x).
\ee
In the approximation considered (52) equation (48) becomes \be
A\f{d\ln\eta}{dx}+2=0\;.
\ee
Let us introduce a new function \be
f(x)=\f{x^3_1}{2}\eta (x)A(x)\;.
\ee
Equation (65) becomes \be
\f{dA}{dx}=1+\epsilon f(x)\;, \ee
and equation (66) takes the following form \be
\f{A}{f}\cdot \f{df}{dx}-\f{dA}{dx}=-2\;.
\ee
From equations (68) and (69) we find \be
A(x)=-\f{(1-\epsilon f)f}{\left (\f{df}{dx}\right )}\;.
\ee
From expression (67) we get \be
\eta (x)=-\f{2\f{df}{dx}}{x^3_1(1-\epsilon f)}\;.
\ee
Substituting (70) and (71) into (47) we  discover \be
U=\f{x^3_1}{2xf},\;\;
V=-\f{x\f{df}{dx}}{f(1-\epsilon f)\left (\f{dz}{dx}\right )^2}\;.
\ee
By using these expressions the determinant $g $ can be written in the
following
form
\be
g=
\f{x^3_1\f{df}{dx}x^4}
{2f^2\left (\f{dz}{dx}\right )^2(1-\epsilon f)}\sin^2\Theta<0\;.
\ee

For fulfilment of  condition (7) it is necessary that expressions
$ \f {df} {dx} $ and $ (1-\epsilon f) $ have opposite signs.
Substituting
(70) into (68) we get \be
\f{d}{dx}\ln
\left |
\f{df}{dx}\right |
-\f{d}{dx}\ln
\left |f(1-\epsilon f)\right |=
\f{1+\epsilon f}{f(1-\epsilon f)}\cdot \f{df}{dx}\;.
\ee
From here we find \be
\f{d}{dx}\ln
\left |
\f{(1-\ep f)\f{df}{dx}}{f^2}\right |=0\;.
\ee
Thus \be
\left |
\f{(1-\ep f)\f{df}{dx}}{f^2}\right |=C_0>0\;.
\ee
Taking into account that the values $ (1-\ep f) $ and $ \f {df} {dx}
$ should have
opposite signs, we obtain \be
\f{df}{dx}=-\f{C_0 f^2}{(1-\ep f)}\;.
\ee
Substituting this expression in (70) we find \be
A(x)=\f{(1-\ep f)^2}{C_0 f},\;\;
A(x_1)=0\;\;\;\mbox{under}\;\;\; f=\f{1}{\ep}\;.
\ee
By taking into account (78) expression (47) for  function $V $
becomes \be
V=\f{C_0 xf}{(1-\ep f)^2\left (\f{dz}{dx}\right )^2}\;.
\ee
Integrating (77) and allowing for (78) we get \be
C_0\cdot (x-x_1)=\f{1}{f}+\ep\ln\ep |f|-\ep\;.
\ee
Relation (80) is obtained in a range of values $x $ determined by
inequalities (60), however, it is also correct  in the area where the
influence
of  a graviton mass can be neglected.

According to (60) the range of $C_0 (x-x_1) $ is confined to limits
\be
-C_0 x_1\lambda_1\leq C_0 (x-x_1)\leq C_0 x_1\lambda_2, 
\ee
if $f$ is positive, it satisfies  inequalities \be
\tilde C\leq f\leq \f{1}{\ep}\;.
\ee
By using (80) and according to (81) we have
$$
\f{1}{f}+\ep\ln\ep f-\ep\leq C_0x_1\lambda_2.
$$
From here it is possible to find $ \tilde C $: \be
\f{1}{\tilde C}+\ep\ln\ep \tilde C-\ep=C_0x_1\lambda_2.
\ee
From expression (83) we can find an approximate value for $ \tilde C
$: \be
\tilde C=\f{1}{C_0x_1\lambda_2}\;.
\ee
For negative values $f$  to  a point $x=x_1 $ corresponds  the value
$ |f | $, determined from the following equation \be
-\f{1}{|f|}+\ep\ln\ep|f|-\ep=0\;.
\ee
From here we get \be
|f|=\f{a}{\ep},\;\;\ln a=\f{1+a}{a}\;.
\ee
According to (81) the following inequality should to be fulfilled
\be
-C_0x_1\lambda_1\leq -\f{1}{|f|}
+\ep\ln\ep |f|-\ep\;.
\ee
From here it is possible to find the lower limit for $ |f | = D $ \be
-C_0x_1\lambda_1=-\f{1}{D}+\ep\ln\ep D-\ep\;.
\ee
From expression (88) we discover an approximate value for $D $ \be
D=\f{1}{C_0x_1\lambda_1}\;.
\ee
It means, that the value of $ |f | $ fulfils the following
inequality:
$$
|f|\geq D=\f{1}{C_0x_1\lambda_1}\;. \eqno{(89')}
$$

Let us establish now the form of dependence of  variable $z $ of $x
$. Substituting (47) in
$ (40 ') $ and allowing for (48), we get \be
A\f{d}{dx}
\left (x\f{dz}{dx}\right )
=2z-x\f{dz}{dx}
\left [1+\ep (x^2-z^2)
-\f{1}{2} \tk x^2 \left (\rho-\f{p}{c^2}\right )\right ]\;.
\ee
In approximation (52) outside of matter equation (90) becomes \be
A\f{d}{dx}
\left (
x\f{dz}{dx}\right )+x\f{dz}{dx}-2z=0\;.
\ee

It is necessary for us to find the regular solution $z (x) $ of
equation (91).
In  equation (91) we shall proceed from  variable $x $ to $f $. By
using
 relation (80)
 \be
x=\f{1}{C_0f}
[C_0x_1 f+1-\ep f+\ep f\ln \ep |f|]\;,
\ee
and allowing for (65), (66) and (83),  equation (91) can be presented
in
the following
form
\be
\f{d^2z}{df^2}+\f{C_0xf+\ep f-1}{C_0f^2 x} \cdot
\f{dz}{df}-\f{2z}{C_0 f^3
x}=0\;.
\ee
By a straightforward substitution we can establish that the
expression
\be
z=\f{x_1}{2}+\f{1}{C_0 f}
[1-\ep f+\ep f\ln\ep |f|] \ee
satisfies  equation (93) up to the value \be
\ep\f{(1-\ep f+\ln\ep |f|)}{C_0^2 xf^3}\;, \ee
which is extremely small in the  neighborhood of the point $x_1 $.
From
expressions (92) and (94) we find \be
z=x-\f{x_1}{2}\;.
\ee
Allowing for this relation and also (79) and (72), we get \be
U=\f{x_1^3}{2xf},\;\;
V=\f{C_0xf}{(1-\ep f)^2}.
\ee
For negative values $f$ the causality condition (51) becomes \be
(2x^2C_0-\ep^2x_1^3)-2\ep x^3_1 |f|-x^3_1\leq 0.
\ee
Inequality (98) is not valid, as it does not fulfil inequality $ (89
') $. Thus, the Principle of  Causality is violated in the region of
negative values of $f$. It means that in the area $x_1(1-\lambda_1)
\leq
x < x_1 $ the solution has no physical sense.
At $x_0 < x_1 (1-\lambda_1) $ the situation arises, when the physical
solution
inside a body $0\leq x\leq x_0 $ cannot be sewed  to the physical
solution in the region $x > x_1 $, as there is an intermediate region
$x_1
(1-\lambda_1) \leq x < x_1 $, in which the solution does not satisfy
the  Causality Principle. It follows from here with necessity that
$x_0\geq
x_1 $. From the physical point of view it is necessary to eliminate
also
the equality $x_0=x_1 $, as the solution inside a body should
continuously pass into
the external solution. Therefore, the variable $f $ takes only
positive
values, and $x_0 $ cannot be less than $x_1 $. For the values from
the
region $x\geq
x_1 (1 +\lambda_2) $  it is possible to omit the terms with a small
parameter $ \ep $ in equations $ (38 ') $ and $ (39 ') $. Thus, we
shall come to the
external Schwarzshild solution \be
z_s=(x-\omega)
\left [
1+\f{b}{2\omega}
\ln\f{x-2\omega}{x}\right ]\;, \ee
\be
V_s=\f{x}{\left (\f{dz}{dx}\right)^2(x-2\omega)}\;,
U_s=\f{x-2\omega}{x}\;.
\ee
Here {\tt "} $ \omega $ {\tt "} and {\tt "} $b $ {\tt "} are some
constants,
which are determined from the condition of sewing  solutions
(96), (97) with the solution (99), (100).
 The function $z $ from (96) is equal to
\be
z=x_1\left (\f{1}{2}+\lambda_2\right )\;,
\ee
 at the point $x=x_1 (1 +\lambda_2)$.
At the same point $z_s $ is equal to
\be
z_s=[x_1(1+\lambda_2)-\omega]
\left [
1+\f{b}{2\omega}\ln
\f{x_1(1+\lambda_2)-2\omega}{x_1(1+\lambda_2)}\right ].
 \ee
From a sewing condition of (101) and (102) we find \be
\omega=\f{x_1}{2},\;\; b=0\;.
\ee
The function $U $ from (97) is equal to \be
U=\f{x^3_1}{2x_1(1+\lambda_2)\tilde C}\;, \ee
at the point $x=x_1 (1 +\lambda_2) $, as $ \tilde C $, according to
(84), is equal to \be
\tilde C=\f{1}{C_0x_1\lambda_2}\;.
\ee
By substituting (105) into (104) we get \be
U=\f{C_0x^3_1\lambda_2}{2(1+\lambda_2)}\;,
\ee
at the same point, with  account for (103), $U_s $ is equal to
\be
U_s=\f{\lambda_2}{1+\lambda_2}\;.
\ee
From a sewing condition  of (106) and (107) we get \be
C_0=\f{2}{x^3_1}\;.
\ee
At the point $x=x_1 (1 +\lambda_2) $ the function $V $ from (97) is
equal to \be
V=C_0x_1(1+\lambda_1)\tilde C\;.
\ee
By substituting the value $ \tilde C $ from (105) into (109) we
obtain \be
V=\f{1+\lambda_2}{\lambda_2}\; ,
\ee
at the same point $V_s $, with  account for (99) and (103), is
equal to \be
V_s=\f{1+\lambda_2}{\lambda_2}\;,
\ee
i.e. the solution for $V $ is sewed  to the solution for
$V_s $.

Let us consider (92) for values $ \ep f $, close to unity \be
f=\f{1}{\ep \left(1+\f{y}{\ep}\right) },\;\;
\f{y}{\ep}\ll 1.
\ee
By substituting this expression into (92) and expanding it over $ \f
{y} {\ep} $, we
obtain \be
y^2=2\ep C_0 (x-x_1).
\ee
Inequality (112) tells us that the value $ (x-x_1) = \delta\ll\ep $,
i.e.
\be
\f{y}{\ep}=\sqrt{2C_0}\cdot \sqrt{\f{x-x_1}{\ep}}\ll 1\;.
\ee
By substituting (113) into (112), and then $f $ into (97), we get the
following expressions
for $U $ and $V $ :
\be
U=\f{x^3_1[\ep +\sqrt{2\ep C_0(x-x_1)}]}{2x}\;,\;\;
V=\f{x [\ep +\sqrt{2\ep C_0(x-x_1)}]}{2\ep(x-x_1)}\;.
\ee
From here we have in the region of  variable $x $, satisfying
inequality (114),
\be
U=\f{\ep x^3_1}{2x},\;\;
V=\f{x}{2(x-x_1)}\;.
\ee

We see, that the presence of  a graviton mass essentially changes the
nature of
 solution in the region close to the gravitational radius. In that
 point, where the
function $V $, according to (116), has a pole, the function $U $ is
different from
zero, whereas in the General Relativity Theory (GRT) it is equal to
zero.
Just by virtue of this circumstance the inevitable gravitational
collapse arises, during which ``black holes'' appear in GRT. In RTG
``black holes'' are impossible.

If we take into account (42), (43), (96) and  neglect the second term
in
(59),  expressions (116) for $U $ and $V $ become \be
U=\left (
\f{GMm}{\hbar c}\right )^2,\;\;
V=\f{1}{2}\cdot \f{r+\f{GM}{c^2}}{r-\f{GM}{c^2}}\;, \ee
which coincides with formulas (18) from paper [1]. Note,
that the residue in
the pole of the function $V $ at $ \ep\not=0 $ is equal to $ \f {GM}
{c^2} $, whereas at
$ \ep=0 $ it is equal to $ \f {2GM} {c^2} $. This is due to the fact
that in  case
$ \ep=0 $ the pole of function $V $ at the point $x=x_1 $ arises
because of the
function $f$,  which  has a pole at this point , whereas at $
\ep\not=0 $ it occurs because of the function $ (1-\ep f) $, which
one, according to
 (92), at the  point $x=x_1 $ comes into zero.

Let us compare now the nature of motion of test bodies in the
effective
Riemannian space with  metric (117) and with the Schwarzshild metric.
We write the interval (21) of the Riemannian space as follows: \be
ds^2=Udt^2-\tilde VdW^2-W^2(d\Theta^2
+\sin^2\Theta d\Phi^2)\;.
\ee
Here $ \tilde V $ is equal to \be
\tilde V (W)=V\left (\f{dr}{dW}\right )^2.
\ee
The motion of a test body takes place along a geodesic line of the
Riemannian space
\be
\f{dv^\mu}{ds}+\Gamma^\mu_{\alpha\beta}v^\alpha v^\beta=0\;, \ee
where
\be
v^\mu=\f{dx^\mu}{ds}\;,
\ee
the four-vector of velocity $v ^\mu $ meets the  following condition: 
\be
g_{\mu\nu}v^\mu v^\nu=1\;.
\ee

Let us consider a radial motion, when \be
v^\Theta =v^\Phi=0\;.
\ee
By taking into account (29), from  equation (120) we find \be
\f{dv^0}{ds}+\f{1}{U}\cdot \f{dU}{dW} v^0v^1=0\;, \ee
where
\be
v^1=\f{dW}{ds}\;.
\ee
From  equation (124) we get \be
\f{d}{dW}\ln (v^0U)=0\;.
\ee
From here we have \be
v^0=\f{dx^0}{ds}=\f{U_0}{U}\;,
\ee
where $U_0 $ is a constant of integrating.

Taking into account (127) we see that condition (122) for radial
motion becomes \be
\f{U^2_0}{U}-1=\tilde V\cdot \left (
\f{dW}{ds}\right )^2.
\ee
If we accept,  the speed of a falling  test body at infinity being 
equal to
 zero, we shall get $U_0=1 $. From (128) is follows \be
\f{dW}{ds}=-\sqrt{\f{1-U}{U\tilde V}}\;.
\ee
Taking into consideration (79), (96), (97) and (108), we have
$$
U=\f{x^3_1}{2xf},\;\;
\tilde V=\f{2xf}{x^3_1(1-\epsilon f)^2}\;.
$$
Substituting these expressions in (129) we get \be
\f{dW}{ds}=-\sqrt{1-U}(1-\ep f)\;.
\ee
By using (108), (112) and (113) in the neighborhood of the point $x_1
$
we have \be
\f{dW}{ds}=-\f{2}{x_1}\sqrt{\f{x-x_1}{\ep x_1}} \ee
Passing from a variable $x $ to $W $, according to (43) and taking
into account (44), we obtain
\be
\f{dW}{ds}=-\f{\hbar c^2}{mGM}
\sqrt{\f{W}{GM}\left (1-\f{2GM}{c^2W}\right )}\;.
\ee
It is apparent from here  that there is a turning point. By
differentiating
(132) on $s$ we get \be
\f{d^2W}{ds^2}=\f{1}{2GM}
\left ( \f{\hbar c^2}{mGM}\right )^2.
\ee
In the turning point the acceleration (133) is rather great, and it
is
positive, i.e. repulsing takes place. By integrating (132) we obtain
\be
W=\f{2GM}{c^2}+
\left ( \f{\hbar c^2}{2mGM}\right )^2\cdot
\f{1}{GM}(s-s_0)^2.
\ee

Formulas (132-134) coincide with the formulas from publication [1].
The presence of the
Planck constant in equation (132) is connected with the wave nature
of matter
formed, in our case, of gravitons having a rest mass. From formula
(134) it is apparent, that the test body can never  intercept the
Schwarzshild sphere. In GRT the situation is rather different. From
the Schwarzshild solution
and expression (129) it follows that the test body will cross
the Schwarzshild sphere and a ``black hole''  will be formed. The
test bodies or light can cross the Schwarzshild sphere only in the
inside
direction, thus
they already can never  leave  the  Schwarzshild sphere.
We shall come to the same result if we proceed to a synchronous
system of freely falling test bodies with the help of transformations
\be
\tau = t+\int dW \left [ \f{\tilde V (1-U)}{U}\right ]^{1/2}.
\ee
\be
R = t+\int dW \left [ \f{\tilde V}{U(1-U)}\right ]^{1/2}.
\ee
In this case interval (118) becomes \be
ds^2=d\tau^2-(1-U)dR^2-W^2
(d\Theta^2+\sin^2\Theta d\Phi^2)\;.
\ee
In such a form  singularities of metric coefficients dissapear both 
for the Schwarzshild solution, when $ \ep=0 $, and for the solution
in our case, when $ \ep\not=0$.

Subtracting from expression (136) expression (135) we get \be
R-\tau =\int
dW\sqrt{\f{U\tilde V}{(1-U)}}\;.
\ee
Differentiating equation (138) over $ \tau $ we discover the
following \be
\f{dW}{d\tau}=-\sqrt{\f{(1-U)}{U\tilde V}}\;.
\ee
Thus, we come to the same initial equation (129), around 
which  the
formulas (132-134) were obtained. Thus, it is abundantly clear that
the transition
to the synchronous falling reference frame does not eliminate the
singularity which
arises due to the presence of a graviton mass, i.e. when $
\ep\not=0$. In
case when
$ \ep=0 $, the Schwarzshild singularity of the metric does not
influence
the motion of a test body both in the initial coordinate
system and in the falling synchronous system. Thus, the falling
particles cross
the Schwarzshild sphere  in the  inside direction only.

Let us calculate
now the propagation time for a light signal from a point $W_0 $ up
to the point
$W_1 =\f {2GM} {c^2} $. For the Schwarzshild solution from expression
$ds^2=0 $
we have
\be
\f{dW}{dt}=-c
\left (
1-\f{2GM}{c^2W}\right )\;.
\ee
By integrating this equation we get \be
W_0-W+\f{2GM}{c^2}\ln
\f{W_0-\f{2GM}{c^2}}{W-\f{2GM}{c^2}}
=c(t-t_0)\;.
\ee
Hence it is apparent that  to achieve the gravitational
radius $W_1=\f {2GM} {c^2}$ in GRT we need an infinite time measured
by a
distant observer clock.
In RTG, as we  have established earlier, the Schwarzshild solution
takes place up to the point $W=W_1 (1 +\lambda_2) $, and therefore
the
time interval to reach this point is equal to
\be
c(t-t_0)=W_0-W_1(1+\lambda_2)+
\f{2GM}{c^2}\ln
\f{W_0-\f{2GM}{c^2}}{\lambda_2\f{2GM}{c^2}}\;.
 \ee
The propagation time of a light ray from the point $W=W_1 (1
+\lambda_2) $
up to the point $W_1 $ can be computed by using  formulas (97) and
(108). In
this interval we have \be
\f{dW}{dt}=-c\f{x^3_1}{2xf}(1-\ep f).
\ee
Hence  after integrating and replacement of a variable we get 
\be
\f{2MG}{c^2}\int\limits^{1/\epsilon}_{f}\f{xdf}{f}=c(t_1-t)\;.
 \ee
According to (84) and (108) the lower limit of integration is equal
to \be
f=\tilde C=\f{x^2_1}{2\lambda_2}\;.
\ee
Integral (144) is easily evaluated and with good accuracy results in 
the following relation: 
\be
c (t_1-t)=W_1\lambda_2+\f{2GM}{c^2}\ln \f{2\lambda_2}{\epsilon}\;.
\ee
On the basis of equations (142) and (146) the time 
needed for a light signal
to pass the distance from the point $W_0 $ up to the point
$W_1=\f{2GM}{c^2}$ is equal to the sum of expressions (142) and (146)
\be
c(t_1-t_0)=W_0-W_1
+\f{2GM}{c^2}\ln
\f{W_0-\f{2GM}{c^2}}{\epsilon \f{GM}{c^2}}\;.
\ee
Thus it is evident that in RTG, as opposed to GRT, the
propagation time of a
light ray up to the  Schwarzshild sphere is finite also if measured
by a distant observer clock. From  formula (147) it is apparent that
the propagation time of
a signal does not sharply increase due to the gravitational field.

From the above it is apparent that in the presence of a
graviton mass
$ \ep\not = 0 $ the solution in RTG differs essentially  from the
Schwarzshild solution
because of the presence of the Schwarzshild sphere  singularity,
which
 cannot be removed by any choice of coordinate system. For this
 reason, as
we have shown above, the physical solution for a static spherically
symmetric
body is possible only in the case, when the point $x_1 $ is inside
the body.
This conclusion is also preserved  for the synchronous coordinate
system, when
the metric coefficients (see  (134)) are functions of time.

Thus, according to RTG as a field
theory of  gravitation, the body of any mass cannot contract
unlimitedly, and therefore the gravitational collapse to form 
a ``black hole'' is impossible. In GRT the energy release at a
spherically symmetric accretion of matter on
a ``black hole'' is not enough, as the falling matter carries
energy into the ``black hole''.

According to RTG, the situation cardinally changes, as at the 
accretion the falling matter  hits the  surface of a body, and
therefore the energy
release is now considerable. The field approach to 
gravitation changes in
essence  our notions which were formed under the
influence of GRT.
In particular, this manifests in the fact that the effective
Riemannian space
which has
arisen due to the  gravitational field, has only simple
topology,
since the gravitational field in  the Minkowski space, as well as
any other
physical field, can be described in a single Galilean coordinate
system.
In GRT the
Riemannian space may have a complicated topology, and it is described
by the atlas of charts.

Further it is noteworthy  that the operation of
a
gravitational field, as well as any other physical field, does not
move the trajectory
 of motion of a test body outside the causality cone of
the Minkowski space. This circumstance  allows  one to compensate the
three-dimensional gravitation force by a force of inertia through
selection of an accelerated
coordinate system.

There is a principal difference between gravitation forces  and
forces of inertia. The force of inertia can always be made equal to
zero by having selected an inertial system of coordinates,
whereas the gravitation  force, which has arisen because of the
presence
of a
gravitational field,  is impossible to be made  equal to zero  by 
selection of a coordinate system, even locally.

If GRT asserts that the gravitation is the consequence  of the
space-time (Riemannian) curvature, then, according to RTG, the
effective Riemannian space-time is a consequent of the presence of a
gravitational field, possessing density of energy-momentum. The
source of it is the energy-momentum tensor  density of the entire
matter,
the gravitational field included.
 The space-time was and is the Minkowski space, and all the
 remaining, including
the gravitation, are physical fields. Just under these notions the
basic physical principles --- the integral  conservation laws of
energy-momentum and angular momentum take place.

The field approach to gravitation with necessity requires the 
introduction of the graviton mass, which, in turn, makes the
gravitational
collapse  impossible
and results in the cyclical development of the homogeneous and
isotropic
Universe. Thus, the homogeneous and isotropic Universe is ``flat'',
and the
existence of ``dark matter'' in the Universe is a forecast [3]. It
follows direct 
from the theory, that present density of matter in
the Universe  should be equal to
\be
\rho (\tau)=\rho_c(\tau)+\rho_g, \ee
where $ \rho_c $ is the critical density, determined by the Hubble
``constant''  $H (\tau) $ and is equal to
\be
\rho_c=\f{3H^2}{8\pi G}\;, \ee
and $ \rho_g $ is determined by the graviton mass $m $ and is equal
to
\be
\rho_g=\f{1}{16\pi G} \left (
\f{mc^2}{\hbar}\right )^2.
\ee

Since critical density $ \rho_c $ many times exceeds 
the observable density of matter in the Universe, then, according to
equation (148), there should be a dark matter in the Universe.

\section*{Appendix A}
In sperical coordinates of the Minkowski space the intervals of the
Minkowski space and of the effective Riemannian spaces look like
$$
d\sigma^2=dt^2-dr^2-r^2
(d\Theta^2+\sin^2\Theta d\Phi^2)\;, \eqno{(A.1)}
$$
$$
ds^2=U(r)dt^2-V(r)dr^2-W^2(r)
(d\Theta^2+\sin^2\Theta d\Phi^2)\;. \eqno{(A.2)}
$$
Let us introduce the velocity vector
$$
v^i=\f{dx^i}{dt},\;\;
v^i=ve^i,\;\; (x^i=r,\Theta,\Phi)\;. \eqno{(A.3)}
$$
where $e^i$ is the unit vector defined by the metric of a spatial
section of the Minkowski space-time
$$
\kappa_{ik}e^ie^k=1\;. \eqno{(A.4)}
$$
In general $ \kappa _ {ik} $ is given as follows
$$
\kappa_{ik}=-\gamma_{ik}+\f{\gamma_{0i}\gamma_{0k}}{\gamma_{00}}\;.
\eqno{(A.5)}
$$
In  case (A.1)
$$
\kappa_{ik}=-\gamma_{ik}. \eqno{(A.6)}
$$
Condition  (A.4) for the metric (A.1)  looks like
$$
(e^1)^2+r^2[(e^2)^2+\sin^2\Theta\cdot
(e^3)^2]=1\;. \eqno{(A.7)}
$$
Let us define four-vector of velocity by the following equation
$$
v^\mu=(1, ve^i)\; \eqno{(A.8)}
$$
and demand that it should be isotropic in the Minkowski space
$$
\gamma_{\mu\nu}v^\mu v^\nu=0\;. \eqno{(A.9)}
$$
By substituting  (A.8)  into  (A.9)  and accounting for  (A.7) we
obtain
$$
v=1\;. \eqno{(A.10)}
$$
Thus, isotropic four-vector $v ^\mu $ is equal to
$$
v^\mu=(1, e^i)\;. \eqno{(A.11)}
$$

As according to the Special Relativity Theory the motion always takes
place inside or on the boundary of the Minkowski causality cone,
the Principle of Causality takes place for the gravitational field
$$
g_{\mu\nu}v^\mu v^\nu\leq 0\;, \eqno{(A.12)}
$$
that is,
$$
U-V(e^1)^2-W^2[(e^2)^2+(e^3)^2\sin^2\Theta]\leq 0\;. \eqno{(A.13)}
$$
By taking into account  (A.7),  expression  (A.13)  can be written as
follows
$$
U-\f{W^2}{r^2}
\left (
V-\f{W^2}{r^2}\right )
(e^1)^2\leq 0\;. \eqno{(A.14)}
$$
Let
$$
V-\f{W^2}{r^2}\geq 0\;. \eqno{(A.15)}
$$
By virtue of an arbitrariness of $0\leq (e^1) ^2\leq 1 $,  inequality
(A.14) will be fulfilled only if
$$
U-\f{W^2}{r^2}\leq 0\;. \eqno{(A.16)}
$$
From inequalities  (A.15)  and  (A.16) it follows that
$$
U\leq V\;. \eqno{(A.17)}
$$
In  case if
$$
V-\f{W^2}{r^2} < 0\;, \eqno{(A.18)}
$$
we shall write inequality (A.14)  in the following form:
$$
U-V-\left (
\f{W^2}{r^2}-V\right )
(1-(e^1)^2)\leq 0\;. \eqno{(A.19)}
$$
By virtue of the arbitrariness of $e^1 $, (A.19) will be satisfied
for any
values of $0\leq (e^1) ^2\leq 1 $ only in case
$$
U\leq V\;.\eqno{(A.20)}
$$

Thus, the  RTG Principle of Causality results in all the cases in the
inequality
$$
U(r)\leq V(r)\;. \eqno{(A.21)}
$$

\end{document}